\documentstyle[epsfig]{article}
\begin{document} \begin{center} {\Large{\bf Emergence of Power Law
in a Market with
Mixed Models.}} \\

M. Ali Saif\footnote{e-mail contact :ali@cms.unipune.ernet.in} 
\\{\em Department of Physics, 
University of Pune, Ganeshkhind, Pune, 411 007,
India}\\ 
 and Prashant M.
Gade\footnote{e-mail contact : gade@unipune.ernet.in} 
\\{\em Center for
Modeling and Simulation, University of Pune, Ganeshkhind, Pune, 411 007,
India}\\ 
\end{center}
                                                                                
\begin{abstract}
 We investigate the problem of wealth
distribution from the viewpoint of asset exchange. Robust nature
of Pareto's law across
economies, ideologies and nations suggests that this could be
an outcome of trading strategies. However, the simple
asset exchange models fail to reproduce this feature.
A yardsale(YS) model in which amount put on the bet is 
a fraction of minimum of the two players
leads to condensation of wealth in hands of some agent while
theft and fraud(TF) model in which the amount to be exchanged
is a fraction of loser's wealth leads to an exponential
distribution of wealth.
We show that if we allow few agents to follow 
a different model than others, {\it i.e.} there are
some agents following TF model while rest follow
YS model, it leads to
distribution with power law tails. Similar effect is observed
when one carries out transactions for a fraction of
one's wealth using TF model and for the rest  YS model is used.
We also observe a power law
tail in wealth distribution  
if we allow the agents to follow either of the models
with some probability.

 PACS:89.65.Gh,64.60.-i

\end{abstract}

\section{Introduction}

Rich get richer and poor get poorer. Worse, rich people do not
seem to be significantly cleverer or more hardworking than
the poorer lot. This has puzzled philosophers and economists
alike all the way from Buddha to Marx. A century ago, an Italian
economist Pareto gave a celebrated empirical law suggesting that it is
just a law of nature that 80\% of the wealth is in 20\% hands.
In fact, on surveying various countries and economies in Europe,
he gave a famous law, now known as Pareto law. It said that the
probability $P(x)$ that an individual has wealth $x$ follows a
power law for large $x$, {\it {i.e.}} $ P(x)\sim x^{-\nu}$.
Distribution of personal wealth and income in countries
as diverse as USA, UK, Japan and India
seem to have a power law tail
\cite{Dragulescu,Fujiwara,Dimatto,Sitabhra,chatyard}.
Since it seems to be independent of the political systems of
those countries, which were widely different, it can be  conjectured
that this distribution is the inherent outcome of economic activities.
In fact, it was recently observed that even in ancient Egyptian society,
wealth distribution could be following the power law\cite{pre}.
In this society, obviously myriad of factors playing role in
modern economies do not exist. Thus it could be conjectured
that there has to be an explanation for the law from simple
and primitive principles of economic activity.
 
One of the striking economic activity that every society, including
the most ancient ones, is capable of,  is `give and take'. Someone
gives you a cup of coffee, you hand him over a dollar. This is
a simple exchange of assets. Let us call it `additive asset exchange'.
However, you do not keep your entire wealth at stake to a coffee
shop owner. But you may keep it at stake in bank. Here you get money
which is proportionate to the money you own. Let us call it
`multiplicative asset exchange'. Researchers have looked
at the models of wealth distribution in presence of additive
and multiplicative asset exchange and results are intriguing.
Somehow,  simple asset exchange models are unable to reproduce
the power law tail which seems to be robust feature across economies.
For multiplicative asset exchange models, in which all agents
start with same wealth, have similar capabilities (none is cleverer
than the other in any sense) the emerging distribution of wealth is
even less equitable than a power law. It turns out that in `free
and fair' trade, one agent (by pure luck, since we have not
assigned extra capabilities to any agent) ends up swallowing
the entire wealth. In another `theft and fraud' rule, we get
an exponential distribution of wealth.
 
What are these models?\cite{Brian Hayes} We consider
two models given by Brian Hayes. In these models,
there is no consumption of wealth nor any production.
In the first model, we assume that everyone knows the value
of everybody else's asset perfectly. This 'free and fair' 
(since nobody is able to conceal true value of his assets) model 
is called Yardsale model\cite{fnn}. 
It is 
the following: There are $N$ individuals
in society and they trade with each other on one-to-one basis.
Everyone is able to value everyone else's assets 
perfectly while trading. Naturally, the amount traded is a fraction
of assets of poorer party.
However, we can have another rule. In this rule, the amount to be exchanged
is a fraction of loser's wealth. Naturally, poorer agents have more
to gain by playing with richer ones and they can do so only by 
deception. Hence it has been named
theft and fraud (TF) model\cite{fn}.
The YS and TF rules can be given as follows.
Let us consider set of $N$ agents with wealth $m_1(0), m_2(0) \ldots
m_N(0)$ at time $T=0$. At each timestep $T=t$ we choose two agents
$i$ and $j$ and their wealths $m_i(t)$ and $m_j(t)$ are updated
as:\\
\begin{eqnarray}
m_i(t+1)=m_i(t)+\Delta m\\
m_j(t+1)=m_j(t)-\Delta m
\end{eqnarray}
where $\Delta m$ is the net wealth exchanged between that two agents.
(Wealth of rest of the agents is unchanged.)
In the YS model, 
$\Delta m=\alpha \min(m_i(t),m_j(t))$.
Whereas, in the TF model the money exchange is fraction of the wealth loser player. 
Then, $\Delta m=\alpha(m_j(t))$ (if $j$ is the loser).
The parameter $\alpha$ is a uniformly distributed 
random number in the interval $\lbrack 0,1\rbrack$.

However, none of these models reproduces the power law
distribution  of wealth found in several societies.
The YS model essentially produces condensation of wealth
in hands of one of the agents. Whereas, the TF model yields an
exponential distribution of assets. 
None of these models reflect the empirically observed distribution  of
wealth \cite{Sinha}.
To mimic these 
observed features of income and wealth distribution, 
several efforts have been made. Some researchers
have applied the techniques of  statistical  physics to the 
economic system.  
They  treat the economic agents as  particles 
in gas. The total wealth is conserved
which is analogous to energy in ideal gas.
Thus the equilibrium probability distribution of money $P(x)$ should 
follow the Boltzmann-Gibbs law $P(x)=c \exp^{-x/T}$.
Here $x$ is money,  $T$ the 
average money per agent and c is constant \cite{Dragulescuand V.M.Yakovenko} .
Chatterjee and Chakrabarti argued that not all the money is
put at stake in market. Every economic agent saves something
for a rainy day.  They 
studied the effect of saving propensities for the agents
\cite{Chakraborti}. Two cases have been studied.
In the first case, all the agents have the same fixed saving factor
\cite{Chakraborti} while in other case
the agents have a quenched 
random distribution of saving factors\cite{Manna}.
The former case yields
the  gaussian distribution of wealth while the later model gives 
a power law distribution of wealth. 
The other model was introduced by Sinha. He assumed that a 
richer player is less likely to be aggressive when bargaining
over a small amount with a poorer player. When this role is 
added to the YS model we can see the exponential and power-law 
distribution of wealth \cite{Sinha}. A similar model was also
introduced by Rodr\'iquez-Achach and Huerta-Quintanilla\cite{Achach}.
Several researchers obtained Pareto-like behavior using  
different approaches such as:
rich people trading with the gross system while poorer
agents continue to have two-party transaction \cite{das}, 
flow of wealth from outside
leading to inelastic 
scattering \cite{slanina}, generalized Lotka Volterra dynamics \cite{solomon} 
and stochastic evolution equation
which incorporate trading as well as random changes in
prices of investments \cite{bouchaud}.  

Here, we would like to argue that entire society playing with a simple model 
is unlikely and unrealistic. It is quite likely that different players
play within different paradigms. Thus it is important to investigate
wealth distribution in societies where  we have mixed models.
In this work, we will study the system from this viewpoint.
In one model, we will let a few players to play by TF
model while  the others to play by YS model.  We will study 
the effect of this mixing on the distribution of wealth at equilibrium.
In the second model, 
the agent 
 will invest some part of his money in YS model and remaining part in 
TF trade system.  

We try yet another system in which players take decision to play
using YS or TF model with certain probability. This could be termed
as a system with multiple strategies since unlike previous case,
players have a choice
to play using either of the models.

We note that  condensation can not be steady state in the above
models. The richest agent, while playing by TF model 
or playing with a player using TF  model could lose his wealth
easily. We observe that few agents using TF strategy would 
significantly change the possible steady state of the system.
All these systems lead to 
a wealth distributions which show a clear power-law tail at
higher values of wealth which is comparable
to realistic situations in some cases.

\section{The Model(s)}
We assume that despite limitations, YS model is the 
most reasonable model of asset exchange. We argue that 
the wealth distributions observed in reality are due
to  perturbations to this model. The perturbations we introduce seem
not only be able to deliver a power law tail in the
wealth distribution of individuals, but we also get a variety of 
exponents as seen in realistic economies. 
We  consider a closed economic system where the total 
amount of money $M$ is conserved and the number of economic agents $N$ 
is fixed. No debt is permitted. Initially, 
we divide the total money $M$ among $N$ agent equally.

We consider three different models. In the
first model, we introduce TF agents
in a pure YS model. In the second model, we presume that all asset exchanges 
are partly TF and partly YS. In the third model, we will let each 
player to choose to use YS or TF model with
certain probability.
 All these variants give us a power-law distribution
of wealth.

(A) We consider a case of one agent playing 
by TF model (say $k$th agent) 
while everyone else plays by YS model.
The transaction will go according to the following scheme.
We choose any two players $i$ and $j$ randomly.
If $i=k$ or $j=k$,  the 
transaction between $i$ and $j$  goes by TF model. 
(We must note that this asymmetry is not an essential 
prerequisite of the results. In the case that, if
$i=k$ or $j=k$, the transaction goes by TS or YF with
equal probability, our results do not change.) 
 If both agents are following YS model, transaction
rule is YS.
We must mention that even one agent playing with TF model 
ruins the possibility of condensation of wealth. Thus, the asymptotic 
distribution is expected to change. We observe that it
changes
significantly even in presence of one agent.

(B) We will let all players to use a part
of their money $\lambda _im_i(t)$ in YS strategy and the
other part $(1-\lambda_i) m_i(t)$ to used it in TF trade. In this case,
the wealth distribution will depend on  distribution of $\lambda_i$s. We study 
two cases of distribution of $\lambda_i$s:
(i)  $\lambda_i$'s have same value for all agents, {\it {i.e.}}
$\lambda_i=\lambda$ for $1\leq i \leq N$.
(ii)  $\lambda_i$'s have a quenched random distribution.
Let us consider this to be uniform distribution over an
interval 
$\lbrack 0,1\rbrack$.

(C) We consider a case in which every agent can trade by either
of models with some probability.
We suppose that  $i$th agent has inclination to trade by YS model with 
probability $p_i$ and by TF model with probability ($1-p_i$). We will let each
agent to choose the value of $p$ from a uniform random number in the interval
$\lbrack 0,1\rbrack$. We will assume the quenched state, where the agents have 
different value of $p$. We conduct the transactions as follows: we select two
agents $i$ and $j$ randomly. The players choose
to trade by TF model  with probability $p_i$ and $p_j$ and by
 YS model with probability
$1-p_i$ and $1-p_j$. If the
two agents chose different model the transaction will occur by YS model.
(Even here, we must mention that asymmetry does not play a significant
role. If we make a rule that transaction will be there {\em {if and
only if}} both agents follow the same model, we get the
same asymptotic distribution.) We observe that the asymptotic
wealth distribution has a power law tail with fairly high value
of exponent.

\section{The Simulation and Result}

In these models, we need to find the asymptotic probability distribution.
We need to employ certain systematic approach to check if the asymptotic
distribution is actually reached. In all these cases we find
that the average wealth of the richest agent as an useful
quantifier. We plot this quantity as a function of time and
have taken the 
saturation of this quantity  as an indicator of the possibility
that the wealth distribution has saturated. (Apart from this 
plot, we have also checked the wealth distribution at 
different timesteps and have checked if it has converged.)

In model A, 
we recorded the wealth of the richest agent. We have introduced only
one TF agent in a sea of YS agents. The average wealth of richest agent
is plotted as a function of time in Fig. 1. 
Here we notice that average wealth of the richest agent saturates to 
{\em {same}} value for all values of $N$. This value is not unity.
Thus one TF agent is able to qualitatively  change the dynamics
of the system. Not only that, all these models show similar characteristics,
though the effect of single TF agent is apparent in a larger system
only after a longer time. 
Let us denote the which wealth of richest agent saturates  
as $t_c$. As expected $t_c$ increases with $N$.
The saturation time $t_c$ scales with $N$ as
$t_c(N)\simeq a N^b$ with $a\simeq 0.90$ and $b\simeq 2.23$. This behavior 
is depicted in Fig. 2.
The   
 $t_c(N)$ gives us an idea about the time needed for the system to attain the steady state.
We study the wealth distributions for $t>t_c(N)$.
In Fig. 3, we show  
the wealth distribution for N=100, for  $t= 10^6$. We average over 
3$\times 10^3$ initial conditions.
The system follows a power-law wealth distribution with  exponent
$\nu \simeq 1.1$.
We checked  the robustness of the distribution at
 various values of $N$, {\it {i. e.}}
$N=300$,  
 at time $t= 10^6$. We again average over 
3$\times 10^3$ initial conditions.  This distribution also follows a power-law 
tail with the same exponent $\nu \simeq 1.1$.
Pareto exponent in this strategy will be $\simeq 0.1$. It is very small
compared to that observed in real economies. 

In  model B, agents use different fractions of their money in
YS and TF models. This could be compared with individuals investing
their money in bonds and stocks. When one has bought stocks, one is 
paid according to performance of the company rather than his own
wealth at that time. Thus it is a realistic situation in modern context.
First, we consider the case in which the fraction $\lambda$ is
a constant, {\it {i.e.}} one  
uses $(1-\lambda)m_i(t)$ with TF model and rest of the money in
YS model. We again study the evolution of wealth of the richest agent as
a function of time. We find that the wealth of richest agent saturates at
certain time and as in previous case, we denote this time by
$t_c(N)$.
In Fig. 4 we have plotted $t_c(N)$ as a function of $N$ for $\lambda=0.999$.
We can see that the saturation time $t_c(N)$ scales with
number of agents as  $t_c(N)\simeq a N^b$ with $a\simeq 513$ and $b\simeq 1.204$.

For $\lambda=1$, we have a pure YS model which leads to
condensation of wealth and $\lambda=0$ we have a TF model which leads to
an exponential distribution.
For $\lambda$ very close to 1 but not exactly 1, 
we observe that
the asymptotic distribution has a power law tail. 
In the general case $0<\lambda<1$, as $\lambda$ increase from $0$ to $1$,
the asymptotic distribution of wealth is observed to go 
from exponential to condensate.
In Fig. 5,  
for $\lambda=0.999$. 
we demonstrate 
the asymptotic wealth distribution. It clearly displays a 
power-law  with  exponent $\simeq 1.5$, 
The simulation was carried out for $N=100$, 
$t= 10^6$ iterations and averaged 
over 3$\times 10^3$ initial conditions.
This model has a tunable parameter $\lambda$ and only
for values close to unity we observe a power law behavior. 

However,
everyone has a different appetite for risk.
We attempt a  model in which $\lambda_i$'s have a 
quenched random distribution.
We consider a
uniform distribution of $\lambda$'s. As in previous case, we find the 
saturation time $t_c$ at which maximum wealth saturates.
In Fig. 6, we have plotted $t_c(N)$ as a function of $N$ for 
this model.
The saturation time $t_c(N)$ scales with
number of agents as  $t_c(N)\simeq a N^b$ with $a\simeq 240$ and $b\simeq 1.561$.
It is interesting to note that
the steady state of wealth distribution has a power-law tail
with $\nu\simeq 2.0$  
as that shown in Fig. 7. However,
in the region corresponding to low wealth, 
the wealth distribution is found to be exponential. The inset of Fig. 7 show 
that behavior. In 
this strategy Pareto exponent is  $\simeq 1.0$.
This strategy seem to be more realistic as compared to model A  
previously discussed and it also gives exponents which are comparable to
realistic case. 

In model C, the transaction is carried on 
depending on the agent's choice who chooses to play with YS or TF
model with certain probability. Now 
some transactions will follow YS model and asset exchange
in rest of the transactions will be decided
by TF model. 
We define the 
saturation time $t_c(N)$ by looking at wealth of richest agent
as done in previous cases. We observe that
$t_c(N)\simeq a N^b $ with
$a\simeq 16.82$ and $b\simeq 1.56$ in this case. (See Fig. 8.)
The asymptotic distribution of  wealth 
shown in Fig. 9. It has a power-law 
tail with exponent $\nu\simeq 3.7$.
In this case Pareto exponent
is $\simeq 2.7$ which is comparable to one observed in societies
like Italy.\cite{italy}  

\begin{tabular}{|l|l|l|l|}
\hline
\multicolumn{4}{|c|}{Table 1: Comparison of Power-law and Lognormal Fits.} \\
\hline
Model & Fit &  $\chi^2/{\rm{Dof}}$ &$R^2$\\ \hline
{Model A} & Lognormal & 3 $\times 10^{-5}$ & 0.4911 \\
 & Power-law & 7 $\times 10^{-7}$& 0.9985\\ \hline
{Model B case a)} & Lognormal & $10^{-5}$ & 0.57 \\
 & Power-law & 5 $\times 10^{-7}$ & 0.9976 \\ \hline
{Model B case b)} & Lognormal & $10^{-5}$& 0.057 \\
 & Power-law &2 $\times 10^{-6}$ & 0.8183\\ \hline
{Model C} & Lognormal & $10^{-3}$& 0.30\\
 & Power-law & $10^{-4}$& 0.87\\
\hline
\end{tabular}\\

We have checked  that the power law is a better fit than lognormal for 
all the cases discussed in the paper in several ways. 
We have checked it visually.  We have checked the goodness of fit by 
finding $\chi^2/DoF$ and $R^2$ for three models by fitting it a 
power law functional form and lognormal fit.  The values are given in Table 1.  
 It is clear that $R^2$ values are higher and very close to unity for
power-law fit
which shows that this
 model is relevant for higher fraction of data 
 and $\chi^2/DoF$ values are lower for a power law fit  which shows
that error is smaller in this fit in all cases.

For models B and C where Pareto exponent is more than one,
we have also plotted the 
 Zipf plot. We order the wealth of agents in descending order
and plot the wealth of $k$th ranked agent as a function of its 
rank $k$. It is
known that if the probability distribution has a tail
 of nature $P(x)\sim x^{-(1+\alpha)}$, the rank distribution
$x_k\sim k^{-1/\alpha}$ where $x_k$ is the  $k$th
largest value in the distribution.  Our Zipf plots
are consistent with this result.

\section{Conclusions}
We point out that having a society in which all agents use the same model
is unrealistic. The agents are likely to use different models.
In this context, we studied YS and TF models. 
For a pure YS model,
 condensation of wealth is observed while a pure TF model leads to
exponential distribution. We have presented three 
different models in which  the above two models are mixed.
In model A, we showed that infinitesimal fraction of TF agents can significantly
alter wealth distribution of society where dominant model is YS. 
If we equate YS model with `honesty', and TF model with `cheating', 
the presence of the other possibility seems to help the society
to have more equitable distribution though attaching these virtues
to these models is debatable\cite{fn}. 
This  mixing gives rise to the wealth distribution with  power-
law behavior with exponent $\nu\simeq1.1$. 
In model B, we considered  each transaction
to be consisting of YS and TF component. It also leads to a power law tail in
wealth distribution. This could be thought as individual investing in
debt market as well as in real estate or bonds where return is 
proportional to performance of the company he invested in.
Here we considered two cases a) Homogeneous agents where they put $\lambda$
fraction of their wealth in  YS model. b) Inhomogeneous agents
putting $\lambda_i$ of his wealth (say $x_i$) in YS model where
$\lambda_i$ have 
 quenched random distribution. In former case, for $\lambda$ close to
one, {\it i.e.} for $\lambda=.999$ we observe 
power-law with exponent $\nu\simeq1.5$. 
In the later case, we observe
that the  wealth distribution has a
 power-law tail with exponent $\nu\simeq 2.0$. Interestingly, we
also recover exponential decay at smaller values of wealth which 
matches with known data about wealth
distribution in United Kingdom and United States \cite{adrian}.   
We  also studied a model in which agents indulge in YS or TF
trading with some probability. It gives a power-law tail
with a larger exponent $\nu \simeq 3.7$.

PMG thanks Dr. S. Sinha for discussions. MAS thanks Govt. of Yemen
for scholarship.

\begin{figure}[htbp]
\label{fig(1)}
\centering
\epsfig{file=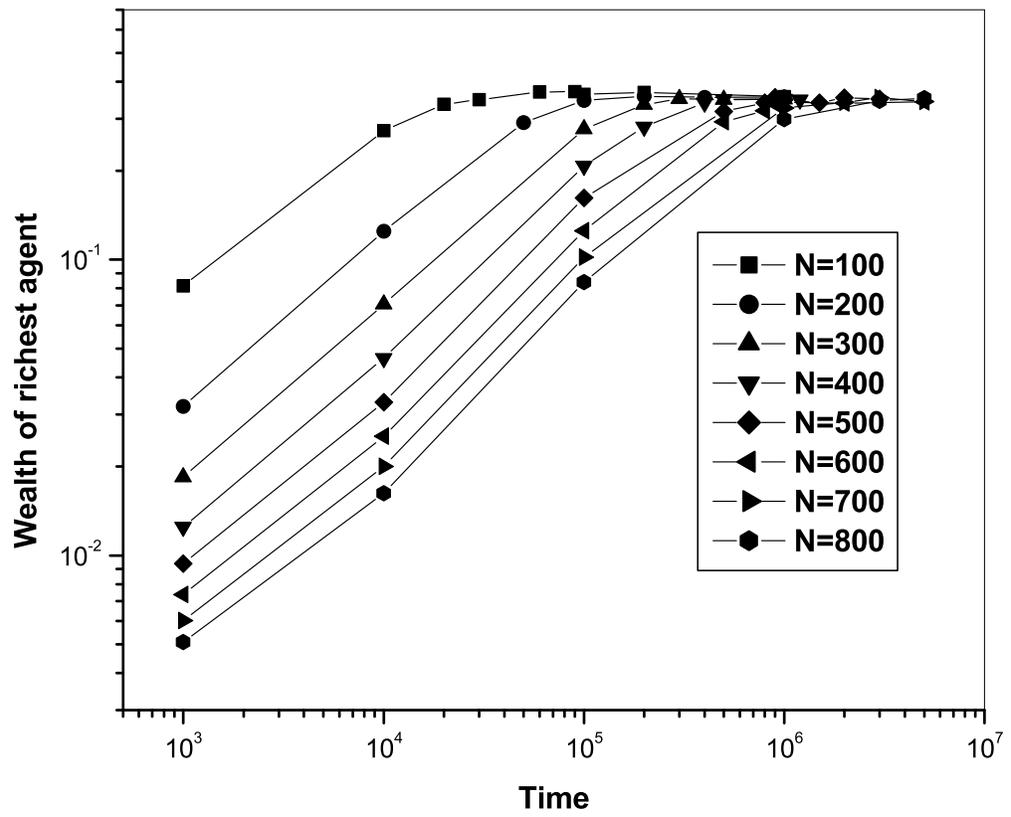,width=15truecm}
\caption{Wealth of the richest agent as function of time step for 
various value of number of agent for model A . We average over
$10^3$ initial conditions.}
\end{figure}

\begin{figure}[htbp]
\label{fig(2)}
\centering
\epsfig{file=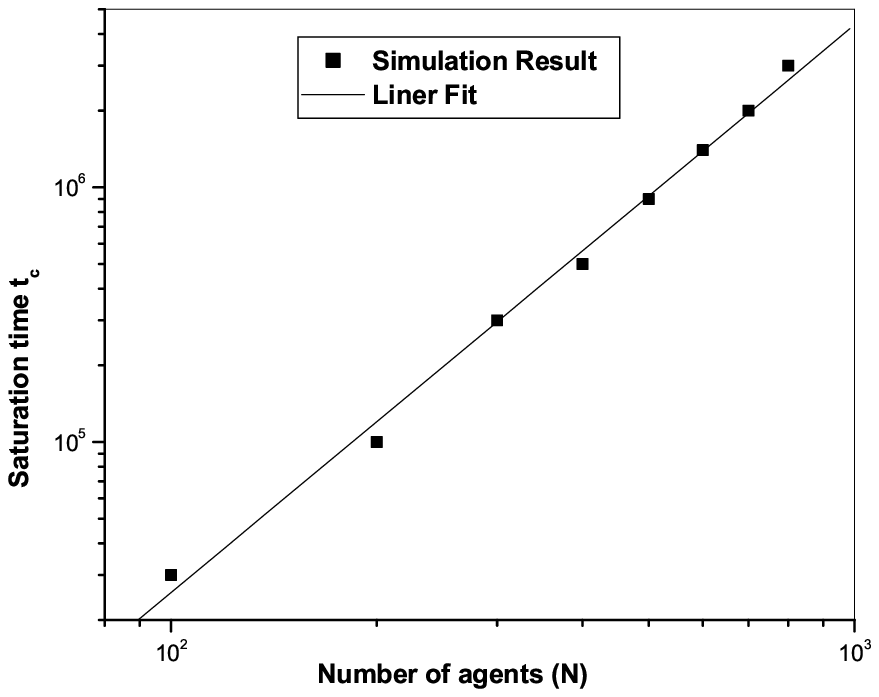,width=6in}
\caption{For model A, we plot  saturation time $t_c$ as
function of number of agent $N$ on logarithmic scale. We average over $10^3$
initial conditions.}
\end{figure}

\begin{figure}[htbp]
\label{fig(3)}
\centering
\epsfig{file=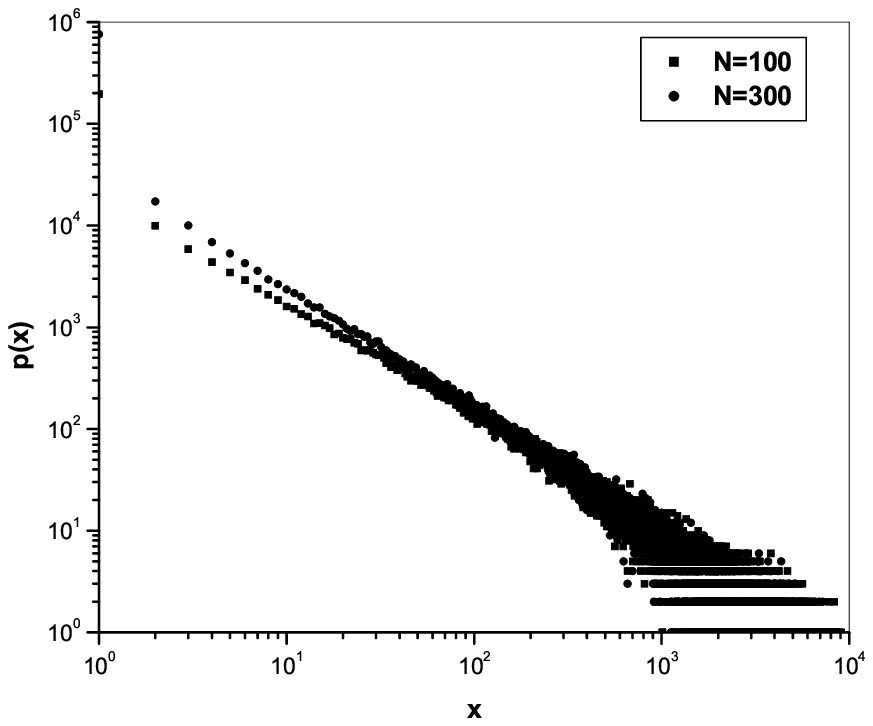,width=6in}
\caption{Asymptotic wealth distribution for model A. We get a power
law with exponent $\nu\simeq1.1$. Simulations are carried out for
 $N=100, 300$. We wait for $10^6 $ 
 transients and  average
over 3$\times 10^3$ initial conditions\cite{fnx}.  }
\end{figure}

\begin{figure}[htbp]
\label{fig(4)}
\centering
\epsfig{file=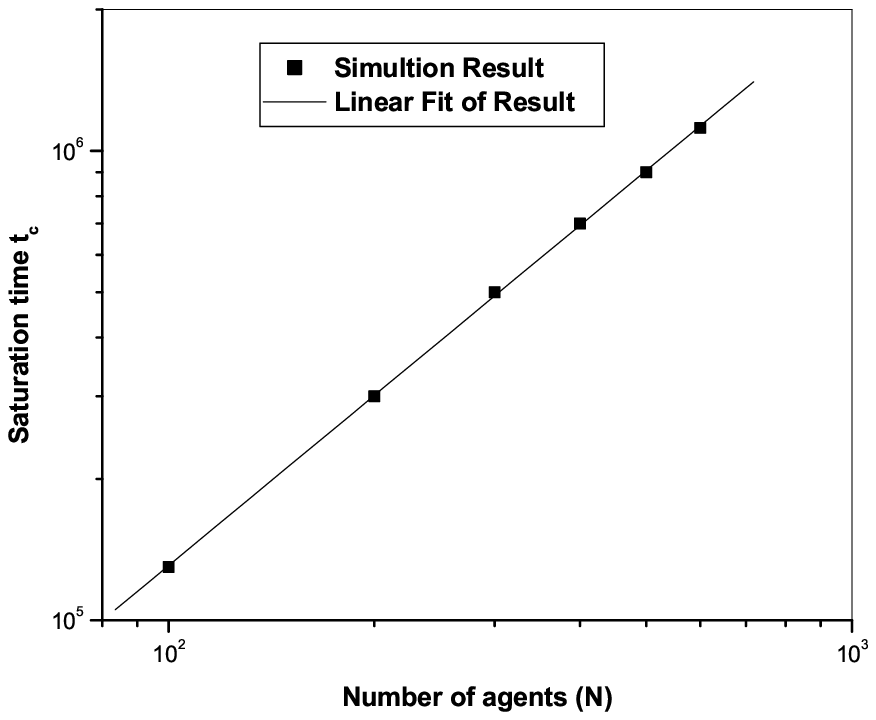,width=6in}
\caption{For model B with homogeneous agents,
we plot saturation time $t_c$ as a function of number $N$ of agents
on logarithmic scale. We average over $10^3$ initial conditions.}
\end{figure}

\begin{figure}[htbp]
\label{fig(5)}
\centering
\epsfig{file=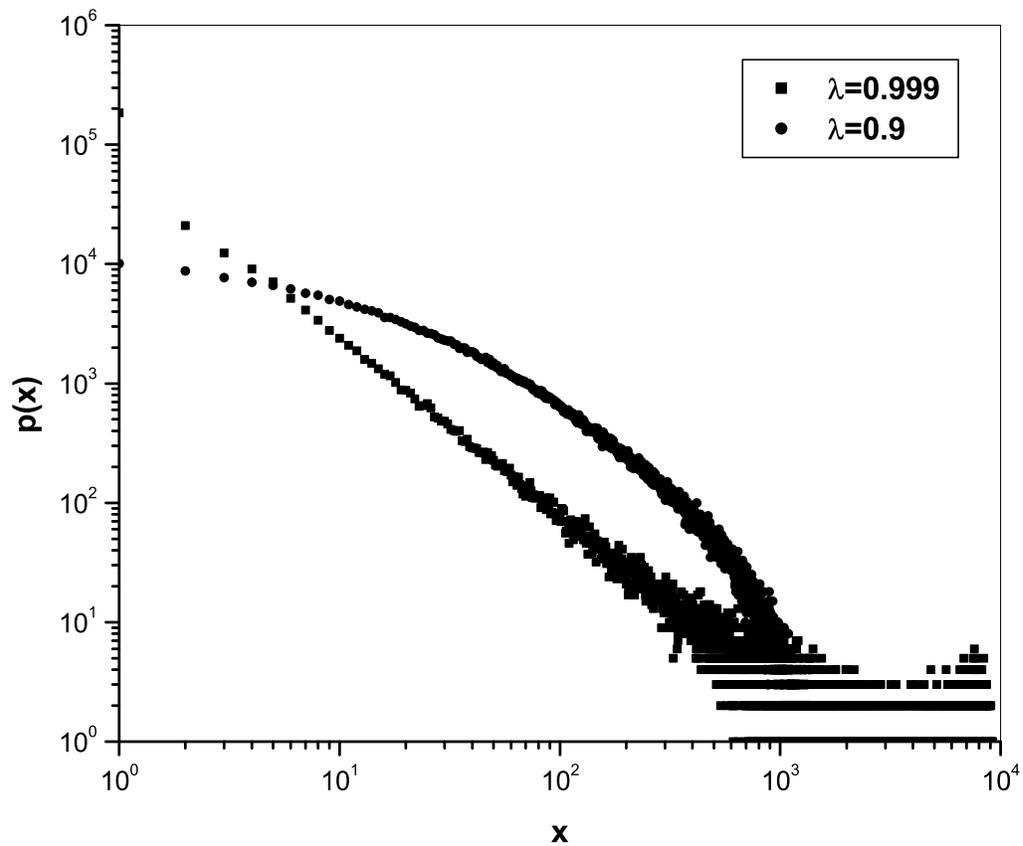,width=6in}
\caption{Asymptotic  wealth distribution for different values of $\lambda$
for model B with homogeneous agents. 
 For $\lambda=0.999$ we get a power-law tail
with exponent $\nu\simeq1.5$. 
Simulations are carried out for $N=100$. We wait for  $10^7$ 
transients and average
over 3$\times 10^3$ initial conditions.}
\end{figure}

\begin{figure}[htbp]
\label{fig(6)}
\centering
\epsfig{file=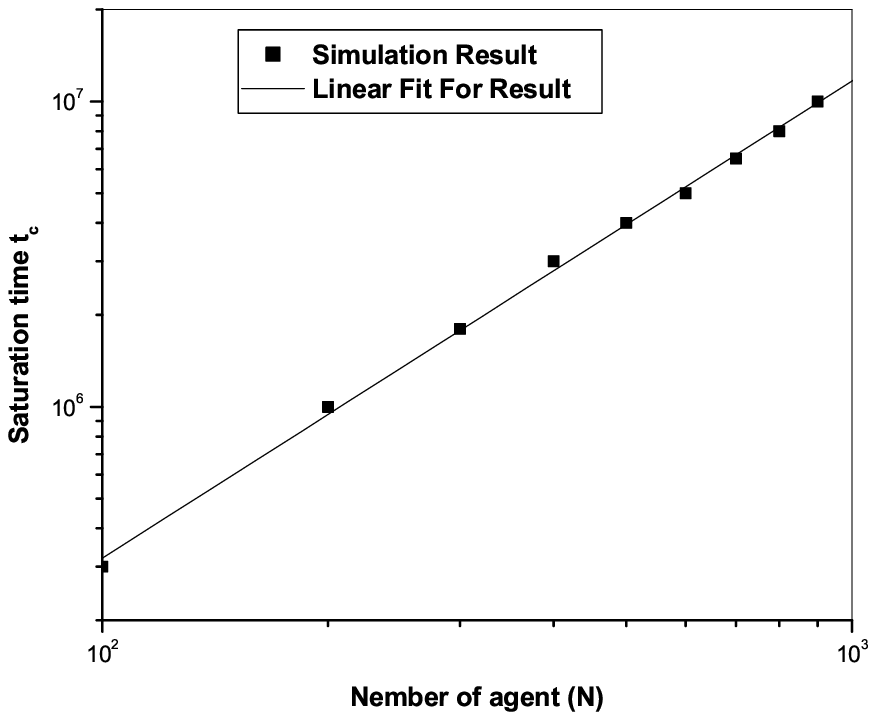,width=6in}
\caption{For model B with inhomogeneous agent, we plot saturation 
time $t_c$ as a function of number 
 of agent $N$ on logarithmic scale. We average over $10^3$ initial conditions.}
\end{figure}

\begin{figure}[htbp]
\label{fig(7)}
\centering
\epsfig{file=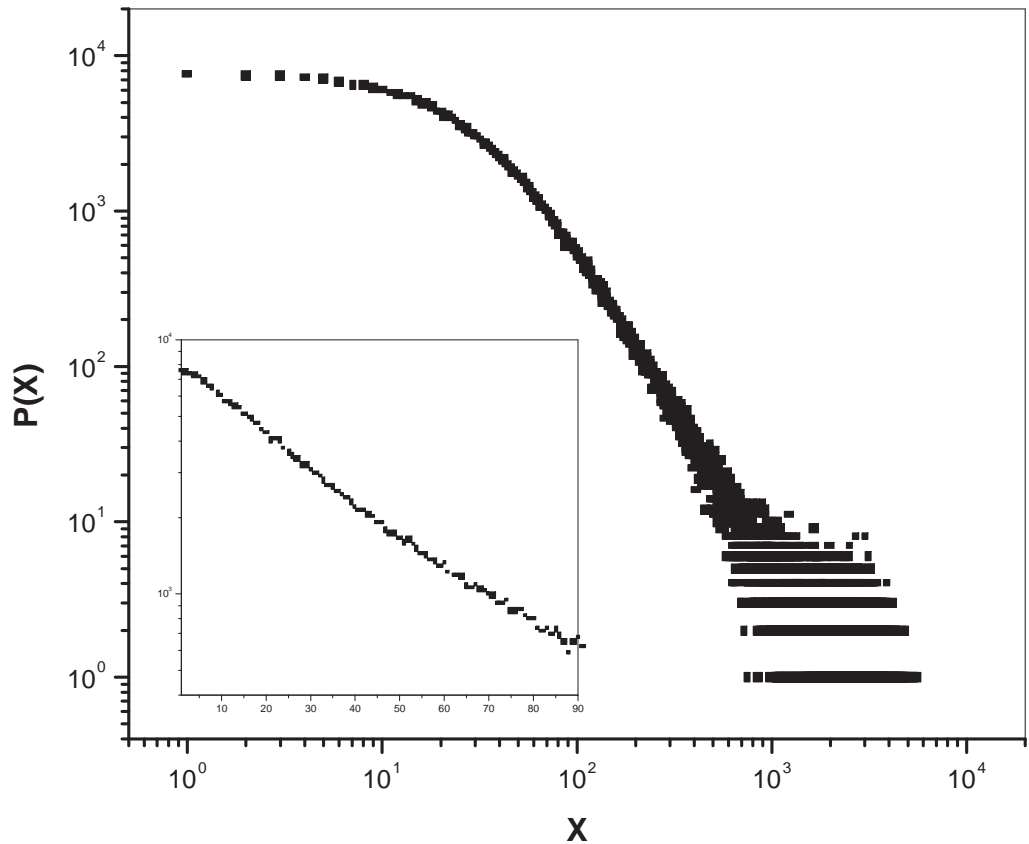,width=6in}
\caption{Asymptotic wealth distribution for model B with inhomogeneous
agent. we get power law tail with exponent $\nu
\simeq 2.0$ at high wealth. Inset: at low wealth, we found an exponential 
wealth distribution. Simulations are carried out for $N=100$. We wait
for $10^7$ transients and average over 3$\times 10^3$ initial conditions. }
\end{figure}

\begin{figure}[htbp]
\label{fig(8)}
\centering
\epsfig{file=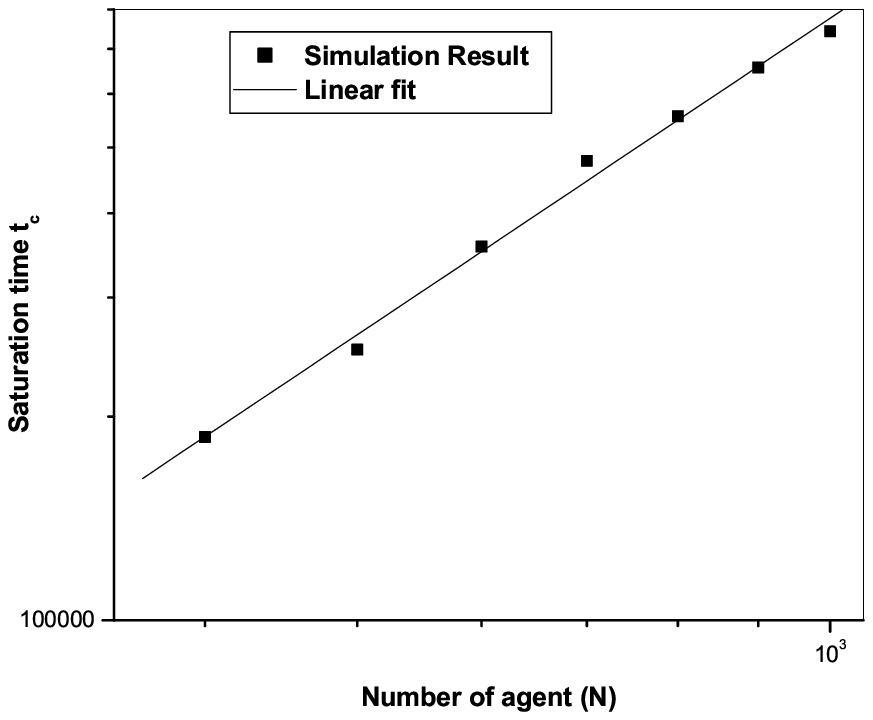,width=6in}
\caption{For model C , we plot saturation
time $t_c$ as a function of number
 of agent $N$ on logarithmic scale. We average over $10^3$ initial conditions.}
\end{figure}

\begin{figure}[htbp]
\label{fig(9)}
\centering
\epsfig{file=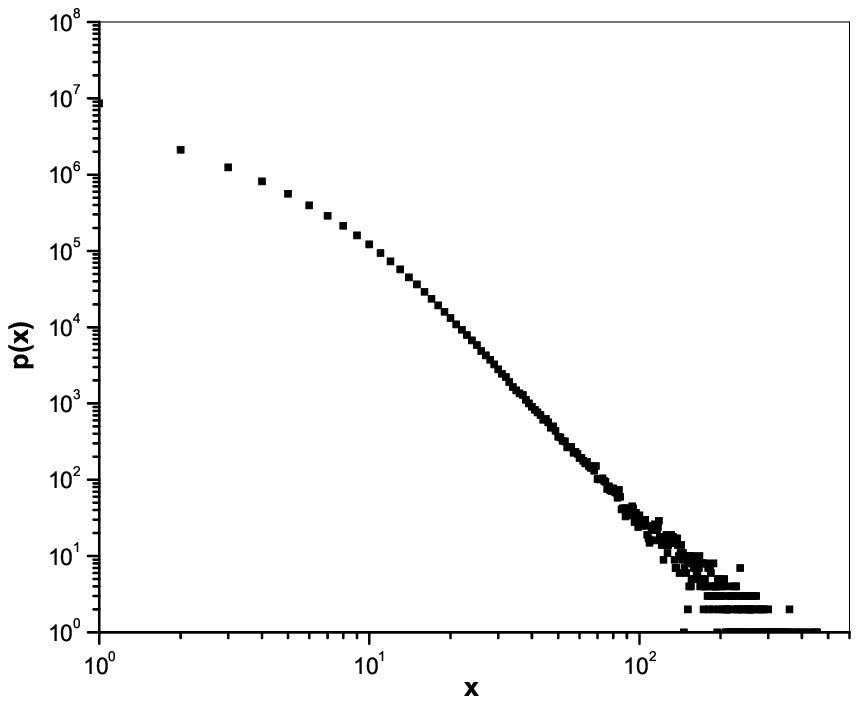,width=6in}
\caption{Asymptotic wealth distribution for model C. We get a power
law with exponent $\nu\simeq 3.7$. Simulations are carried out for
 $N=5000$. We wait for $10^7 $
 transients and  average
over 3$\times 10^3$ initial conditions.  }
\end{figure}


\begin{thebibliography}{99}
\bibitem{fnn}
This is obviously unrealistic
assumption since if there is no arbitrage or bargain, there
would be no reason to trade.
\bibitem{fn}
However, too much should not be read in this choice of names.
Since these names have been adopted by other researchers as well,
we stick to them.
\bibitem{fnx}
The left-most
 point is an artifact of numerics. What we plot is not a probability 
 density but essentially a histogram with really fine bins. Since
 there is a singularity near zero, first point has a lot of weight.
\end{thebibliography}
\end{document}